\documentclass[preprint,aps,showpacs]{revtex4}
\usepackage{graphicx}
\usepackage{dcolumn}
\usepackage{bm}
 
\newcommand{\be}{\begin{equation}}
\newcommand{\ee}{\end{equation}}

 \newcommand{\eqa}{\begin{eqnarray}}
\newcommand{\eeq}{\end{eqnarray}}  
\begin{document}
\title{Dynamics of Fully Nonlinear Drift Wave-Zonal Flow Turbulence System in Plasmas}
\vspace{3cm}
\author{P. K. Shukla}
\email{ps@tp4.rub.de}
\affiliation{Institut f\"ur Theoretische Physik IV, Fakult\"at f\"ur Physik und
Astronomie, Ruhr-Universit\"at Bochum, D-44780 Bochum, Germany}
\author{Dastgeer Shaikh} 
\email{dastgeer.shaikh@uah.edu}
\affiliation{Department of Physics and Center for Space Plasma and Aeronomic Research,
The University of Alabama in Huntsville, Huntsville. Alabama, 35899}
\vspace{2.in}
\received{3 October 2009}
\accepted{8 October 2009} 
\begin{abstract}
We present numerical simulations of fully nonlinear drift wave-zonal
flow (DW-ZF) turbulence systems in a nonuniform magnetoplasma. In our
model, the drift wave (DW) dynamics is pseudo-three-dimensional
(pseudo-3D) and accounts for self-interactions among finite amplitude
DWs and their coupling to the two-dimensional (2D) large amplitude
zonal flows (ZFs). The dynamics of the 2D ZFs in the presence of the
Reynolds stress of the pseudo-3D DWs is governed by the driven Euler
equation. Numerical simulations of the fully nonlinear coupled DW-ZF
equations reveal that shortscale DW turbulence leads to nonlinear
saturated dipolar vortices, whereas the ZF sets in spontaneously and
is dominated by a monopolar vortex structure. The ZFs are found to
suppress the cross-field turbulent particle transport. The present
results provide a better model for understanding the coexistence of
short- and large-scale coherent structures, as well as associated
subdued cross-field particle transport in magnetically confined fusion
plasmas.
\end{abstract}
\pacs{52.35.Kt,52.35.Mw,52.25.Fi,52.35.Ra}
\maketitle
\newpage


It is widely recognized that the presence of large scale sheared flows
\cite{r1,r2,r3} [also referred to as convective cells (CCs) or zonal
flows (ZFs)] is detrimental to regulating the cross-field turbulent
transport in magnetically confined fusion plasmas.The ZF is
characterized by poloidally and toroidally symmetric structure with
radial variation, and the relative zonal flow potential fluctuation
(in comparison with $T_e/e$, where $T_e$ is the electron temperature
and $e$ is the magnitude of the electron charge) is much smaller than
the relative zonal flow density perturbation (in comparison with the
equilibrium plasma number density $n_0$). The large scale Zonal jets also 
occur in various planetary atmosphere, where they are nonlinearly generated 
by the Rossby waves \cite{r4a,r4b}, and influence the atmospheric wind 
circulation \cite{r4c,r4d}.

In magnetically confined fusion plasmas, there exist free energy
sources in the form of density, temperature,and magnetic field
inhomogeneities, which are responsible for exciting the low-frequency
(in comparison with the ion gyrofrequency), short scale (of the order
of the ion gyroradius or the ion sound gyroradius) DW-like
fluctuations \cite{r5a,r5b,r5c}. The linearly growing drift modes interact
among themselves and attain large amplitudes in due course of time. 
The Reynolds stress of finite amplitude DWs, in turn, nonlinearly generate
convective cells (CCs) and sheared flows/ZFs \cite{r6,r7,r8,r9,r10,r11,r12},
via three-wave decay and modulational instabilities \cite{r7}, respectively.  
There are recent review articles presenting the status of theoretical and
simulation works \cite{r12}, as well as experimental observations
\cite{r13} concerning the dynamics of DW-ZF turbulence system.  Specifically,
some numerical simulations \cite{r12} lend support to the experimental
observation that the DW turbulence and transport levels are reduced in the
presence of the sheared flows/ZFs.

Recently, Guo {\it et al.} \cite{r14} used the governing equations of
Ref. \cite{r7} for the DW-CC turbulence system to investigate the
radial spreading of the DW-ZF turbulence via soliton formation.
However, the authors of Ref. \cite{r14} completely neglected
self-interactions among drift waves and zonal flows, which are very
important in the study of nonlinearly coupled finite amplitude drift
and zonal flow disturbances in nonuniform magnetoplasmas.

In this Letter, we present simulation results of fully nonlinear DW-ZF
turbulence systems, which exhibit the coexistence of drift dipolar
vortices and a radially symmetric monopolar zonal flow vortex. The
effect of the latter on the cross-field turbulent transport is
examined.  Our investigation is based on the governing equations for
the DW-ZF turbulence systems that incorporate the Hasegawa-Mima (HM)
self-interaction nonlinearity \cite{r15} in the nonlinear dynamics of
the DW which are nonlinearly exciting CCs/ZFs.  Furthermore, we also
account for nonlinear interactions among the CCs/ZFs and obtain the
driven Euler equation for the dynamics of finite amplitude
CCs/ZFs. The generalization of the governing equations for fully
nonlinear DW-ZF turbulence systems is rather essential for the
investigation of the formation of coherent nonlinear structures that
control the transport properties and confinement of tokamak plasmas.


We consider a nonuniform magnetoplasma in an external magnetic field $\hat {\bf z} B_0$,
where $\hat {\bf z}$ is the unit vector along the $z-$ axis in a Cartesian coordinate 
system and $B_0$ is the strength of the homogeneous magnetic field. The density gradient 
$\partial n_0/\partial x$ is along the $x-$ axis. In the presence of the finite amplitude 
low-frequency (in comparison with the ion gyrofrequency $\omega_{ci} =eB_0/m_i c$, where 
$m_i$ is the ion mass and $c$ is the speed of light in vacuum) electrostatic DWs and ZFs, 
the perpendicular (to $\hat {\bf z}$ component of the electron and ion fluid velocities \cite{r8} 
are, respectively,
\begin{equation}
{\bf u}_{e\perp}^d \approx \frac{c}{B_0} \hat {\bf z} \times \nabla \phi 
-\frac{c}{B_0 n_e} \hat {\bf z} \times \nabla (n_e T_e) \equiv {\bf u}_{EB}^d + {\bf u}_{De}^d,
\end{equation}
\begin{equation}
{\bf u}_{e\perp}^z \approx (c/B_0) \hat {\bf z} \times \nabla \psi \equiv {\bf u}_{EB}^z, 
\end{equation}
\begin{eqnarray}
{\bf u}_{i\perp}^d \approx {\bf u}_{EB}^d + {\bf u}_{Di}^d 
- \frac{c}{B_0 \omega_{ci}} \left(\frac{\partial}{\partial t}
+ \nu_{in} - 0.3\nu_{ii}\rho_i^2 \nabla_\perp^2 +   {\bf u}_{EB}^d\cdot \nabla
+ {\bf u}_{Di}^d\cdot \nabla\right)\nabla_\perp \phi \\ \nonumber
-\frac{c}{B_0\omega_{ci}}\left[({\bf u}_{EB}^d\cdot \nabla)\nabla_\perp \psi
+({\bf u}_{EB}^z\cdot \nabla)\nabla_\perp \phi\right],
\end{eqnarray}
and
\begin{equation}
{\bf u}_{i\perp}^z \approx {\bf u}_{EB}^z -\frac{c}{B_0\omega_{ci}}
\left[\left(\frac{\partial }{\partial t} + \nu_{in}- 0.3 \nu_{ii}\rho_i^2\nabla_\perp^2\right)\nabla_\perp \psi
+\left<({\bf u}_{EB}^d\cdot\nabla)\nabla_\perp \phi\right>\right],
\end{equation}
where the superscripts $d$ and $z$ represents quantities associated with the DWs and ZFs, respectively, 
$\phi$ and $\psi$ are the electrostatic potentials of the DWs and ZFs, respectively,  
$n_e$ and $n_i$ are the electron and ion number densities, respectively, ${\bf u}_{Di}^d 
=(c/eB_0n_i)\hat {\bf z} \times \nabla (n_i T_i)$ is the ion diamagnetic drift velocity, 
$T_i$ is the ion temperature, $\nu_{in} (\nu_{ii}$ is the ion-neutral (ion-ion) collision frequency, 
$\rho_i =V_{Ti}/\omega_{ci}$ is the ion gyro-thermal radius, and $V_{Ti}$ is the ion thermal speed,   
We stress that the self-interaction nonlinearities of the DWs and ZFs are retained in the fluid velocities 
(3) and (4), respectively. The angular brackets denote averaging over one period of the DWs. 

Assuming that $\left|(\partial/\partial t) + {\bf u}_{EB}^d \cdot \nabla\right| \ll \nu_{en}$,
where $\nu_{en}$ is the electron-neutral collision frequency, we obtain from the parallel 
(to $\hat {\bf z}$ component) of the electron momentum equation the magnetic field-aligned electron 
fluid velocity $u_{ez}^d \approx \left(1/m_e \nu_{en}\right) \partial 
\left(e \phi -T_e n_{e1}^d/n_0\right)/\partial z$, where $n_{e1} = (n_e - n_0) \ll n_0$. 
We can now insert $u_{ez}^d$ into the electron continuity equation to obtain
\begin{equation}
\left[\frac{\partial}{\partial t} + \frac{V_{Te}^2}{\nu_{en}} \frac{\partial^2}{\partial z^2} 
+ \left({\bf u}_{EB}^d+ {\bf u}_{EB}^z\right)\cdot \nabla \right]n_{e1}^d
+ {\bf u}_{EB}^d\cdot \nabla n_0 + \frac{n_0 e}{m_e \nu_{en}} \frac{\partial^2\phi}{\partial z^2} =0, 
\end{equation}
where $V_{Te} =(T_e/m_e)^{1/2}$ is the electron thermal speed and $m_e$ is the electron mass. 
Furthermore, substituting for the ion fluid velocity from (3) into the ion continuity equation we have
\begin{eqnarray}
\left[\frac{\partial}{\partial t} + \left({\bf u}_{EB}^d+ {\bf u}_{EB}^z\right)\cdot \nabla \right]n_{i1}^d
+ {\bf u}_{EB}^d \cdot \nabla (n_0 + n_{i1}^z) \\ \nonumber
- \frac{n_0 c}{B_0 \omega_{ci}} \left[ \left(\frac{\partial}{\partial t}
+ \nu_{in} - 0.3 \nu_{ii} \rho_i^2 \nabla_\perp^2 + {\bf u}_{EB}^d \cdot \nabla \right)\nabla_\perp^2 \phi 
+ \nabla \cdot ({{\bf u}_{Di}}^d \cdot \nabla)\nabla_\perp \phi \right] \\ \nonumber
-\frac{n_0c}{B_0 \omega_{ci}}\left[({\bf u}_{EB}^d\cdot \nabla)\nabla_\perp^2 \psi
+({\bf u}_{EB}^z\cdot \nabla)\nabla_\perp^2 \phi\right] =0,
\end{eqnarray}
where the magnetic field-aligned ion dynamics has been ignored, thereby isolating the ion sound waves
from our system. The ion density perturbation associated with the ZFs is 
$n_{i1}^z =\left(n_0c/B_0 \omega_{ci}\right) \nabla_\perp^2 \psi$.

Equations (5) and (6), which govern the dynamics of collisional drift waves \cite{r16} 
in the presence of zonal flows, are closed by assuming $n_{e1}^d \approx n_{i1}^d \equiv n_1$, which is 
a valid approximation in plasmas with $\omega_{pi}^2 \gg \omega_{ci}^2$, where $\omega_{pi}$ is the ion 
plasma frequency. In the linear limit, without the ZFs, Eqs. (5) and (6) yield the DW frequency
$\omega_k = - k_y c_s \rho_s/ L_n (1+k_\perp^2 \rho_s^2)$ and the growth rate $\gamma_k (\ll \omega_k)$, 
which are much larger than the damping rate $\nu_{in} + 0.3 \nu_{ii} k_\perp^2 \rho_i^2$. The
growth rate is $\gamma_k =\nu_{en}\omega_k^2 k_\perp^2/\omega_{LH}^2 k_z^2(1+k_\perp^2 \rho_s^2)$, 
where $c_s =(T_e/m_i)^{1/2}$ is the ion sound speed, $\rho_s =c_s/\omega_{ci}$ 
is the sound gyroradius, $\omega_{LH} =(\omega_{ce}\omega_{ci})^{1/2}$ is the lower-
hybrid resonance frequency, $\omega_{ce}=eB_0/m_ec$ is the electron gyrofrequency, 
$L_n =\left(\partial {\rm ln } n_0/\partial x\right)^{-1}$ is the scale-length of the density gradient, 
and ${\bf k} = {\bf k}_\perp + \hat {\bf z} k_z$ is the wave vector.   

The equation for the ZFs is obtained by inserting (2) and (4) into the electron and ion
continuity equations, and inserting the resultant equations into the Poisson equation,
obtaining the driven [by the DW Reynolds stress; the last term in the left-hand side of Eq. (7)] 
damped (by the ion-neutral collision and ion-gyroviscosity effects) ZF equation
\begin{equation}
\left(\frac{\partial}{\partial t}+ \nu_{in} -0.3 \nu_{ii} \rho_i^2 \nabla_\perp^2 
+ {\bf u}_{EB}^z \cdot \nabla\right)\nabla_\perp^2 \psi
+ \left<({\bf u}_{EB}^d \cdot \nabla) \nabla_\perp^2\phi\right> =0.  
\end{equation}

For the collisionless DWs, we assume that $|(\partial \phi/\partial t)+ ({\bf u}_{EB}^d+{\bf u}_{EB}^z)
\cdot \nabla \phi| \ll (V_{Te}^2/\nu_{en}) \nabla_\perp^2 \phi$ and $\hat {\bf z} \times \nabla n_0 \cdot \nabla
\phi \ll (\omega_{ce}/\nu_{en}) n_0 |\partial^2\phi/\partial z^2|$, and obtain from (5) the Boltzmann law
for the electron number density perturbation $n_{e1}^2 =n_0 e\phi/T_e$. The latter can be inserted into 
(6) by assuming that $n_{i1}^d =n_{e1}^d$, so that we have fully nonlinear equation for the DWs in the 
presence of ZFs 
\begin{eqnarray}
\frac{\partial \phi}{\partial t}-  \frac{c_s \rho_s}{L_n} \frac{\partial \phi}{\partial y}
- \rho_s^2 \left[\frac{\partial}{\partial t} + \nu_{in} - 0.3 \nu_{ii} \rho_i^2 \nabla_\perp^2 
+ \frac{c}{B_0} \left(1+ \sigma\right) (\hat {\bf z} \times \nabla \phi) \cdot \nabla \right] 
\nabla_\perp^2 \phi \\ \nonumber   
+ \frac{c}{B_0} (\hat{z} \times \nabla \psi) \cdot \nabla \left( \phi - \rho_s^2 \nabla_\perp^2 \phi\right) =0,
\end{eqnarray}
where $\sigma =T_i/T_e$.

We normalize the time and space variables by $\omega_{ci^{-1}}$ and $\rho_s$, as well as $\phi$ and $\psi$ 
by $T_e$, and the collision frequencies by $\omega_{ci}$. In the normalized units, we can rewrite (7)
and (8) as, respectively,

\begin{equation}
\left[\frac{\partial}{\partial t}+ \frac{\nu_{in}}{\omega_{ci}} -0.3 \frac{\nu_{ii}}{\omega_{ci}}
 \sigma \nabla_\perp^2 +   (\hat {\bf z} \times \nabla \psi) \cdot \nabla\right]\nabla_\perp^2 \psi
+   \left< (\hat {\bf z} \times \nabla \phi \cdot \nabla) \nabla_\perp^2\phi\right> =0,  
\end{equation}
and
\begin{eqnarray}
\frac{\partial \phi}{\partial t} -  \frac{\rho_s}{L_n} \frac{\partial \phi}{\partial y}
- \left[\frac{\partial}{\partial t} + \frac{\nu_{in}}{\omega_{ci}} - 0.3 \frac{\nu_{ii}}{\omega_{ci}}
\sigma \nabla_\perp^2 + (1+ \sigma) (\hat {\bf z} \times \nabla \phi) \cdot \nabla \right] 
\nabla_\perp^2 \phi 
\\ \nonumber   
+ (\hat{z} \times \nabla \psi) \cdot \nabla \left( \phi -  \nabla_\perp^2 \phi \right) =0.
\end{eqnarray}


We have developed a 2D code to numerically integrate the system of
equations (9) and (10), which describe the self-consistent evolution
of the DW-ZF turbulence systems.  We have chosen $\nu_{in}/\omega_{ci}
=0.1$, $\nu_{ii}/\omega_{ci}=0.01$, $\sigma =0.1$, and $\rho_s/L_n
=0.01$. Numerical descritization employs the spatial derivative in
Fourier spectral space, while time is descritized using time-split
integration algorithm, as prescribed in Ref. \cite{r17}. Periodic
boundary conditions are used along the $x$ and $y$ directions.  A
fixed time integration step is used. The conservation of energy
\cite{r18} is used to check the numerical accuracy and validity of our
numerical code during the nonlinear evolution of the small scale drift
wave fluctuations and zonal flows.  We also make sure that the initial
fluctuations are isotropic and do not influence any anisotropic flow
during the evolution. Anisotropic flows in the evolution can, however,
be generated from a $k_y=0$ mode that is excited as a result of the
nonlinear interactions between the ZFs and small scale DW
turbulence. The ZF and DW fields are initialized with a small
amplitude and uniform isotropic random spectral distribution of
Fourier modes in a 2D computational domain.  These fields further
evolve through Eqs. (9) and (10) under the influence of nonlinear
interactions.  Intrinsically, the set of Eqs.  (9) and (10) possesses
parametrically unstable modes involving short scale drift waves and
zonal flows.  In the early phase of simulations, we obtain the growth
of small scale DWs.  We have carried out two characteristically
distinct sets of simulations by switching on and off the
self-interaction terms. This enables us to gain considerable insight
into the physics of generation of zonal flows and associated transport
level in the coupled DW-ZF turbulence systems.

\begin{figure}
\begin{center}
\includegraphics[width=16cm]{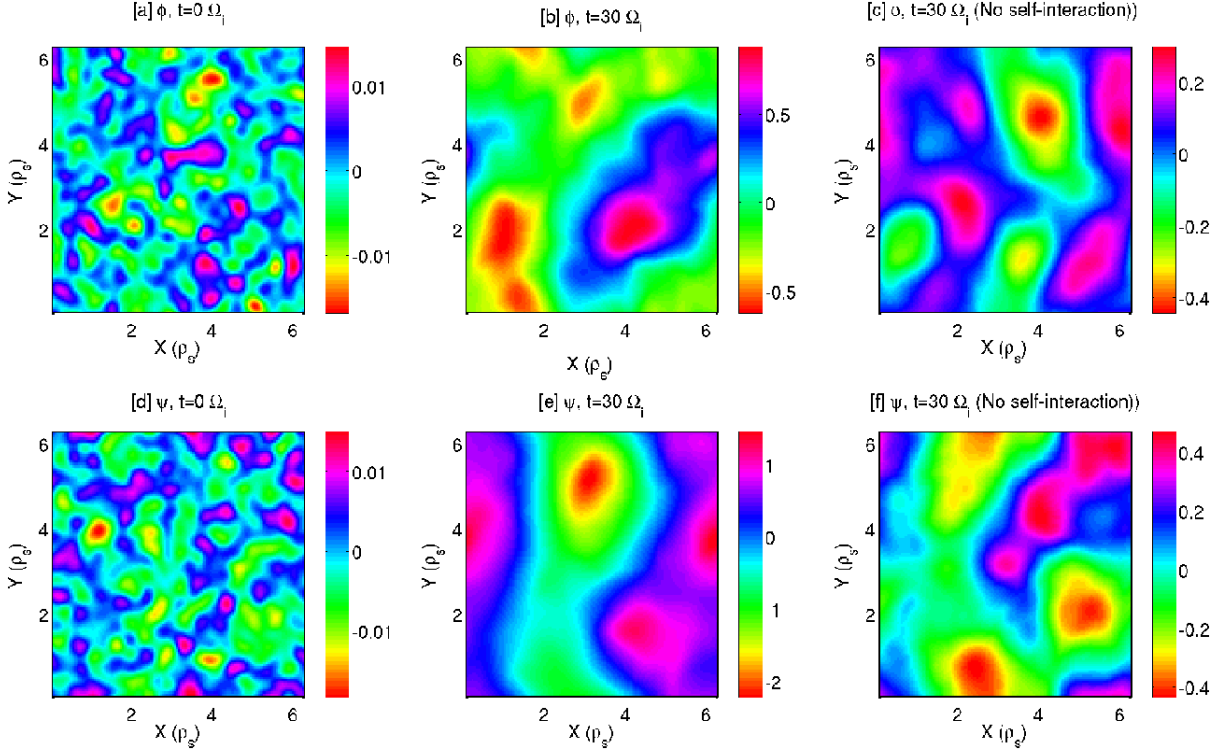}
\end{center}
\caption{Evolution of mode structures in our coupled DW-ZF turbulence model 
from an initial random distribution. In the presence of self-interaction
terms, zonal flows are enhanced and quench the DW turbulence more
efficiently. Numerical resolution is $256^2$, box size is $2\pi
\times 2\pi$.}
\label{energy}
\end{figure}

\begin{figure}[t]
\includegraphics[width=12cm]{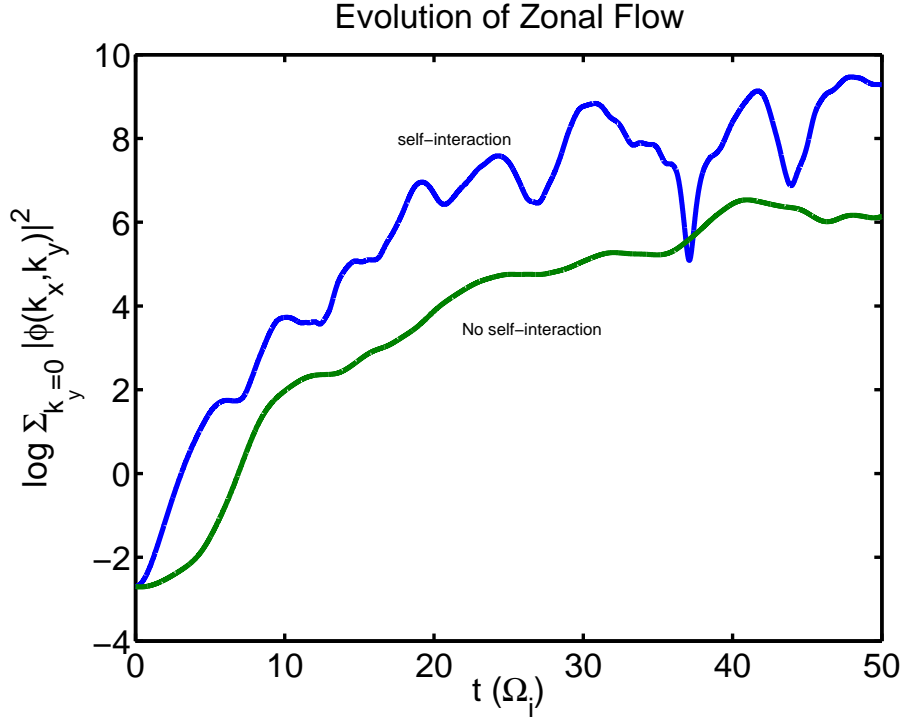}
\caption{The self-consistent generation of zonal flows is shown. In
the presence of the self-interaction nonlinearity, zonal flows are
generated rapidly and their saturated level is also enhanced when
compared with the evolution without the self-interaction nonlinearity.}
\label{spectra}
\end{figure}

\begin{figure}[t]
\includegraphics[width=12cm, height=7cm]{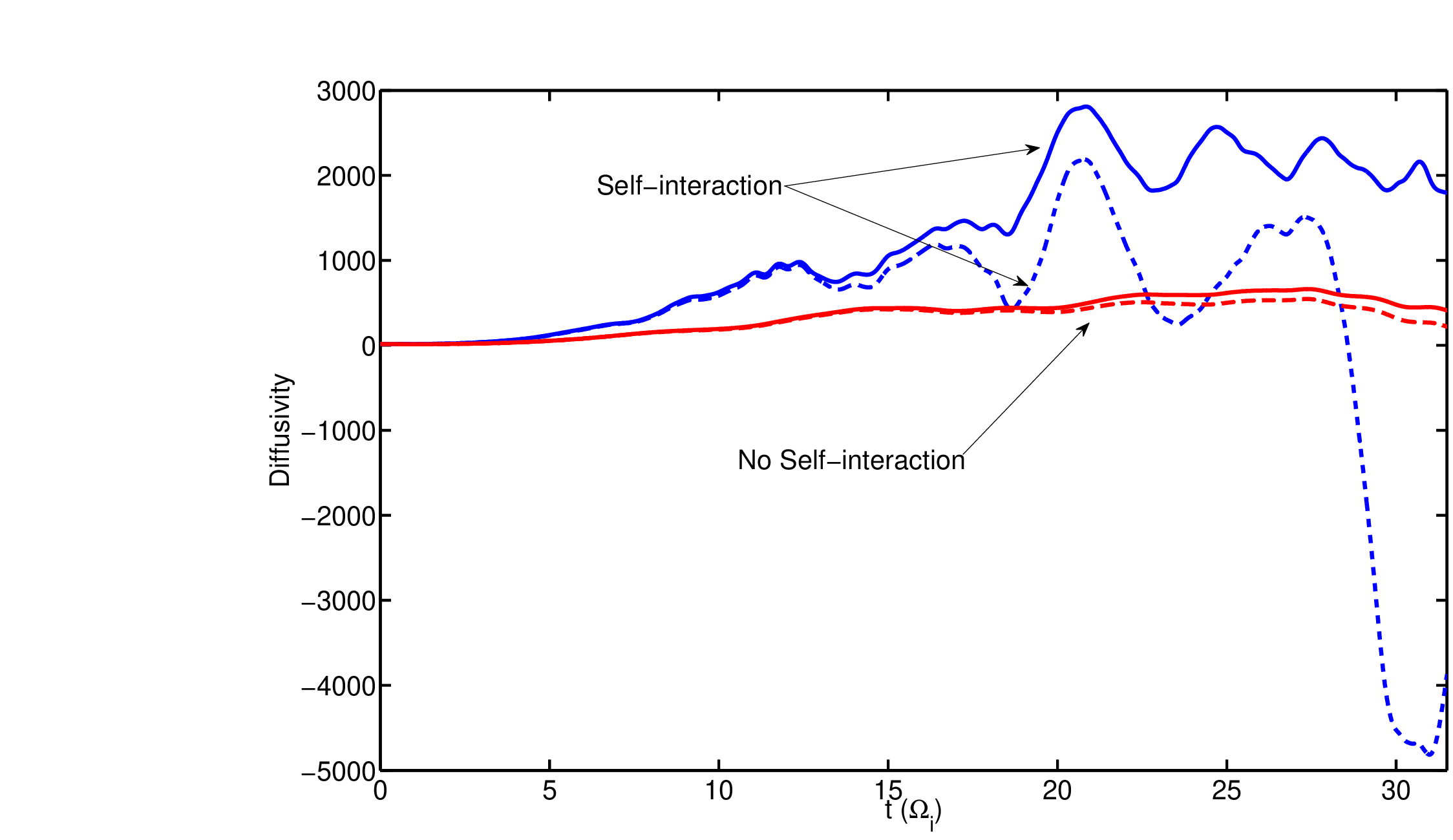}
\caption{Evolution of the cross-field diffusivity in the presence (blue curve), as well
  as in the absence (red) of the self-interaction term. The cross-field diffusivity with
  and without zonal flows is shown by the dashed and
  solid curves, respectively.  Clearly, the presence of the self-interaction term
  enhances zonal flows, which dramatically reduce the cross-field diffusivity.}
\label{spectra2}
\end{figure}

To gain insight into the characteristics nonlinear interactions in our
coupled DW-ZF turbulence model, we examine the
Hasegawa-Mima-Wakatani (HMW) model \cite{r15,r16} that describes the
electrostatic drift waves in an inhomogeneous magnetoplasma.  First,
the ion polarization drift nonlinearity in the HMW model,
viz. $\hat{z}\times \nabla \phi \cdot \nabla \nabla^2 \phi$, signifies
the self-interaction Reynolds stress that plays a critical role in the
formation of the ZFs \cite{r7}. This nonlinearity is basically
responsible for the generations of the ZFs. Secondly, since in
the collisionless DW dynamics, the electron density perturbation follows 
a Boltzmann law due to the rapid thermalization of electrons along $\hat
{\bf z}$, the nonlinearity $\hat{z}\times \nabla \psi \cdot \nabla
\phi$ comes from the cross-coupling of the ZF's ${\bf E}_\perp^z
\times \hat {\bf z}$ particle motion with the drift wave density fluctuation in 
our model. The role of this nonlinearity has traditionally been identified
as a source of suppressing the intensity of the nonlinear flows in the
DW turbulence \cite{r18}. Nevertheless, the presence of the
linear inhomogenous background can modify the nonlinear mode couplings
in a subtle manner. Our objective here is to understand the latter in
the context of the coupled DW-ZF turbulence system.  The initially
isotropic and homogeneous spectral distribution associated with
potential fluctuations, as described above, evolve dynamically
following the set of Eqs. (9) and (10). 

The small amplitude initial drift wave fluctuations are subject to the
{\em modulational instability} on account of their nonlinear coupling with ZFs. 
The parametrically unstable fluctuations grow rapidly during the early phase of
the evolution. The instability eventually saturates via the nonlinear mode
couplings in which the DW Reynolds stress, in concert with other nonlinearities 
in Eqs. (9) and (10), play a critical role.  The mode couplings during the 
nonlinear phase of the evolution leads to the formation of non-symmetric zonal 
flow structures.  This is shown in Fig. 1.  Our simulations exhibit that the
self-interaction terms not only suppress the modulational instability on a
rapid timescales, but they also regulate the generation of the ZFs (see, Figs 1 b, c,
e, f). The extent and amplitude of the ZFs in Figs. 1(b) and
(e) [with the self-interaction] are larger than that of (c) and (f)
[without the self-interaction]. The final (i.e. steady-state)
structures, nonetheless, show the formation of a predominantly dipolar
vortex in the DW fluctuations, while the ZFs are dominated
by a large-scale monopolar-vortex motions.  It is noteworthy that the absence
of the self-interaction contaminates the flows with more small scale
structures (See Figs. 1 c and f).

We next investigate the quantitative evolution of the ZFs, which is
depicted in Fig. 2. The spectral transfer of energy in the ZF
is estimated from $\sum_k |\phi(k_x, k_y=0)|^2$. The latter
describes a pile up of energy in the $k_y=0$ mode that is summed up
over the entire turbulent spectrum.  We find from our simulations that
the presence of the self-interaction nonlinearity rapidly suppresses
the linear phase of the modulational instability of the DWs. Hence, the linear 
phase during the evolution terminates rapidly when compared with the 
no-self-interaction case. Alternatively, the modulational growth rate is enhanced, 
and it saturates on a rapid timescales. Consequently, the presence of the
self-interactions gives rise to an enhanced level of ZFs, as shown in Fig. 2.

A direct consequence of the enhanced ZFs is to markedly suppress the cross-field 
turbulent transport, because the sheared flows (in the poloidal direction) associated
with the ZFs tear apart the DW fluctuations/eddies, and thereby keeping their amplitudes 
low.  We have computed the cross-field diffusivity in our simulations 
by using the ion fluxes involving the Boltzmann electron density 
perturbation and the linear ion polarization drift velocity associated with
the nonthermal DWs. The cross-field ion diffusivity reads
\be
D = {D_B} \sum_{\bf k} \frac{k_y^2 \rho_s^2  |\phi_k|^2}{(1+k_\perp^2 \rho_s^2)},
\ee
where $D_B =cT_e/eB_0$ is the Bohm diffusion coefficient and $\phi_k$
is the spectral potential distribution of the DWs.  Note that the ion thermal 
diffusivity, as defined above, is subdued in the presence of the ZFs, since 
the latter eventually suppress the DW turbulence so that the steady state  
cross-field transport level is reduced.  Consistent with this scenario, 
we find, from our simulations, that the onset of the ZFs quenches the 
cross-field turbulent transport, as shown by the dashed-curve in Fig. 3. 
Furthermore, due to the vanishing poloidal wavenumber of the ZFs, the
sheared flows do not cause any cross-field turbulent transport in magnetized
plasmas.


In summary, the most notable point that emerges from our simulations of the 
coupled DW-ZF turbulence system is the importance and significance of the 
self-interaction nonlinearity in modeling the low-frequency DW turbulence 
that is believed to be a critical source for heat and energy losses. 
A most realistic and accurate understanding of the latter is, therefore, 
essential for the building and the performance of the next generation
controlled thermonuclear fusion reactors, such as the ITER.  In the work, 
we have, for the first time, brought about the importance of self-interaction
processes and their role with regard to the cross-field turbulent transport 
in high-temperature plasmas of thermonuclear fusion devices (e.g. tokamaks).  
For our purposes, we used a new set of nonlinear equations for the 
coupled DW-ZF turbulence system, which is a generalization of Ref. 7, 
by including self-interactions among DWs which drives finite amplitude ZFs.
Numerical simulations of the newly obtained nonlinear equations reveal 
that the coupled DW-ZF turbulence system evolves in the form of short-scale
drift dipolar vortices and a large scale monopolar zonal flow structure.
The simultaneous presence of the dipolar and monopolar 
vortices is responsible for a subdued cross-field turbulent
transport in a magnetically confined fusion plasma.
   


This research was partially supported by the Deutsche Forschungsgemeinschaft
through the project SH21/3-1 of the Research Unit 1048.

\end{document}